# Development of a Hardware-in-the-loop Testbed for Laboratory Performance Verification of Flexible Building Equipment in Typical Commercial Buildings

**Zhelun Chen, PhD**
*Associate Member ASHRAE*
**L. James Lo, PhD**
*Member ASHRAE*
**Amanda Pertzborn, PhD**
*Associate Member ASHRAE*
**Gabriel Grajewski**

**Jin Wen, PhD**
*Member ASHRAE*
**Zheng O'Neill, PhD, PE**
*Fellow ASHRAE*
**Caleb Calfa**
*Student Member ASHRAE*
**Yicheng Li**

**Steven T. Bushby**
*Fellow ASHRAE*
**W. Vance Payne, PhD**
*Member ASHRAE*
**Yangyang Fu, PhD**

**Zhiyao Yang, PhD**
*Member ASHRAE*

**ABSTRACT**

*The goals of reducing energy costs, shifting electricity peaks, increasing the use of renewable energy, and enhancing the stability of the electric grid can be met in part by fully exploiting the energy flexibility potential of buildings and building equipment. The development of strategies that exploit these flexibilities could be facilitated by publicly available high-resolution datasets illustrating how control of HVAC systems in commercial buildings can be used in different climate zones to shape the energy use profile of a building for grid needs. This article presents the development and integration of a Hardware-In-the-Loop Flexible load Testbed (HILFT) that integrates physical HVAC systems with a simulated building model and simulated occupants with the goal of generating datasets to verify load flexibility of typical commercial buildings. Compared to simulation-only experiments, the hardware-in-the-loop approach captures the dynamics of the physical systems while also allowing efficient testing of various boundary conditions. The HILFT integration in this article is achieved through the co-simulation among various software environments including LabVIEW, MATLAB, and EnergyPlus. Although theoretically viable, such integration has encountered many real-world challenges, such as: 1) how to design the overall data infrastructure to ensure effective, robust, and efficient integration; 2) how to avoid closed-loop hunting between simulated and emulated variables; 3) how to quantify system response times and minimize system delays; and 4) how to assess the overall integration quality. Lessons-learned using the examples of an AHU-VAV system, an air-source heat pump system, and a water-source heat pump system are presented.*

**Jin Wen**, **L. James Lo**, and **Zhelun Chen** are Professor, Associate Professor, Research Scientist, **Gabriel Grajewski** and **Yicheng Li** are PhD students in the Department of Civil, Architectural & Environmental Engineering at Drexel University, Philadelphia PA, USA, respectively. **Zheng O'Neill**, **Yangyang Fu**, **Zhiyao Yang**, and **Caleb Calfa** are Associate Professor, Research Engineer, Research Engineer, and PhD student in the Department of Mechanical Engineering at Texas A&M University, College Station, TX, USA, respectively. **Steven T. Bushby**, **Amanda Pertzborn**, and **W. Vance Payne** are the leader of and supervisory electronics engineer in the Mechanical Systems and Controls Group, Mechanical Engineer in the Mechanical Systems and Controls Group, and Mechanical Engineer in the HVAC&R Equipment Performance Group, in the Building Energy and Environment Division of the Engineering Laboratory at the National Institute of Standards and Technology, Gaithersburg, MD, USA, respectively.

## INTRODUCTION

With the increase of renewable energy and the development of the smart grid, it is important to improve the flexibility of both the supply-side and demand-side of the electric grid to meet the needs of power generation, transmission, distribution, and dispatch. As one of the main users of the electric grid, the building sector (including both residential and commercial) plays an important role in the realization of a flexible grid. Fully exploiting the energy flexibility potential of buildings and building equipment can effectively achieve the goals of reducing energy costs, shifting electricity peaks, increasing the renewable energy use in the sector, and enhancing the stability of the grid.

Research has shown that buildings and building equipment can provide flexible electrical loads to improve grid resilience and overall efficiency, and to reduce peak demand (EPRI 2009). However, existing research has mostly focused on developing enabling technologies such as HVAC control strategies (Li et al. 2016). The time-varying values of flexible loads and energy efficiency measures (i.e., practices that lead to energy reduction with insignificant impact on service levels) are not well-studied. These time-varying values are affected by both the time-varying value of electricity and the end-use load shape, energy savings shape (i.e., the difference between electricity demand in the baseline condition and the post-installation electricity demand of the energy efficiency measure), and control. As a result, considerations of the impact of energy systems on peak demand reduction and grid resilience have been limited (Frick et al. 2017). Based on a recent literature review performed by the Lawrence Berkeley National Laboratory, there is a lack of publicly available high-resolution datasets on end-use load and energy savings shape (Frick et al. 2017). Limited existing data are concentrated regionally and are largely in the residential sector; there is a lack of hourly load and energy savings shape data that focus on HVAC systems for commercial buildings, especially under different climate zones and grid programs. There has been little non-residential end-use data research since the End Use Load and Consumer Assessment Program (ELCAP) study (Frick et al. 2017), and existing data often suffer from poor accuracy and low transferability since HVAC load data are weather-sensitive. There is a need for load flexibility data that considers: 1) equipment load characteristics under a wide range of operating conditions and control strategies; 2) interactions between HVAC systems, occupants, and the grid; 3) the effect of building-grid integration on occupant behavior and thermal comfort, which is a market barrier for the adoption of grid-interactive efficient buildings (GEBs). Without these data, it is difficult for utilities to design effective policies and programs, and for manufacturers to design equipment to optimize load flexibility.

To efficiently generate high fidelity equipment performance data, a hardware-in-the-loop (HIL) approach is adopted in this study. Compared to simulation-only experiments, the HIL approach captures the dynamics of the physical systems while also allowing efficient testing of various boundary conditions. This article presents the development and integration of a HIL Flexible load Testbed (HILFT) that integrates physical HVAC systems with a simulated building and simulated occupants with the goal of generating datasets to verify load flexibility of typical commercial buildings. These datasets will be generated from a variable-air-volume system with chillers and an ice tank in the Intelligent Building Agents Laboratory (IBAL) at the National Institute of Standards and Technology (NIST) (Pertzborn 2016; Pertzborn and Veronica 2018), a two-stage air-source heat pump (ASHP) at NIST, and a water-source heat pump (WSHP) at Texas A&M University (TAMU) (Calfa et al. 2022). The testbeds corresponding to these systems are referred to as IBAL, ASHP, and WSHP HILFTs in this paper, respectively.

This paper is organized as follows: Section 2 gives an overview of the HILFTs. Section 3 summarizes the lessons learned from the HIL integration. Section 4 provides conclusions as well as future works.

## OVERVIEW OF THE HARDWARE-IN-THE-LOOP TESTBEDS

In this section, the framework of the HILFT as well as its major components are briefly introduced. Figure 1 depicts the overall framework of the HILFT. A HILFT mainly includes three parts, i.e., the virtual building model, GEB control model, and the hardware testbed. The virtual building model further includes the zone load, occupant comfort & behavior, and airflow models. Details about the HILFT components are described below.

## Virtual Building Model

**Zone Load Model.** The zone load model for this study is adapted from existing commercial prototype building models (U.S. DOE n.d.) and enhanced commercial prototype building models (Pang et al. 2020) developed in EnergyPlus (Crawley et al. 2001). Based on our industry advisory board's suggestions, prototype buildings from five of the ASHRAE climate zones (Baechler et al. 2015), 3A, 4A, 5A, 2B, and 3B, are selected to provide a diversity of weather conditions to which a significant portion of the commercial building stock is exposed nationally. Each EnergyPlus model is exported as a Functional Mockup Unit (FMU) using the EnergyPlusToFMU package (LBNL 2020), and then imported to the MATLAB Simulink environment (The MathWorks 2020) for co-simulation with other models and communication with the equipment.

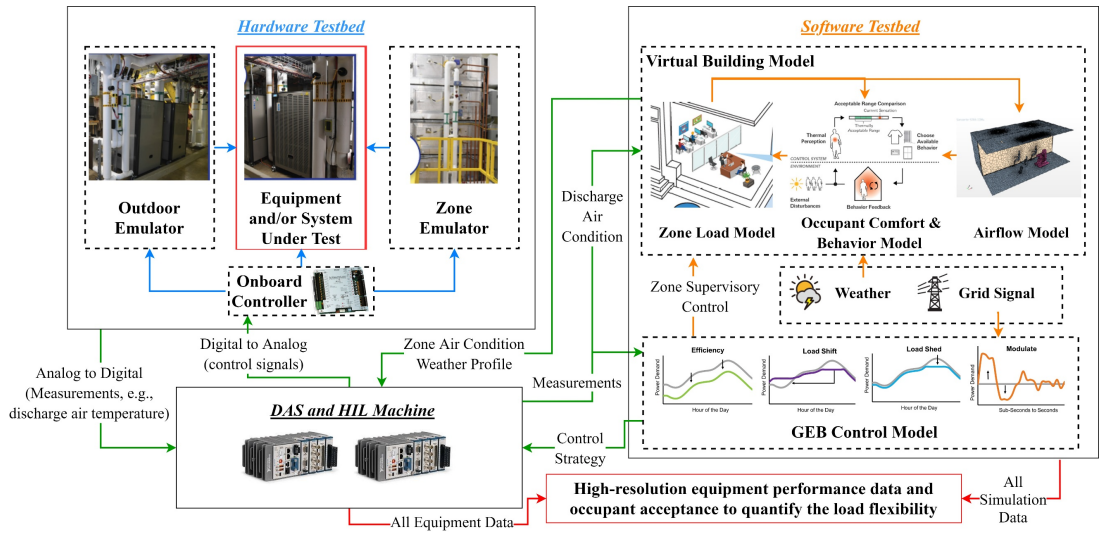

**Figure 1** Framework of the hardware-in-the-loop flexible load testbed and associated data flow schema

**Occupant Comfort & Behavior Model.** Occupant behavior has a significant impact on building load and energy use. An agent-based occupant comfort & behavior model (Langevin et al. 2015) is adapted in this study. This model forecasts the occupants' thermal behaviors, which are fed into the zone load model to determine their impact on the zone thermal load. In this study, occupants have the option (with given probabilities) to turn personal heaters/fans on/off, adjust thermostats, take clothes on/off, have a hot/cold drink, and walk to reduce discomfort. This model is integrated into the Simulink environment as a Level-2 MATLAB S-Functions block.

**Airflow Model.** An indoor airflow model is employed to better estimate the thermal comfort condition surrounding the occupants. The indoor environment is simulated in real-time using a computationally efficient artificial neural network (ANN) model that is trained using computational fluid dynamics (CFD) based simulation data (Zhang et al. 2021). This model uses the discharge air condition measured from the equipment, the zone surface temperatures from the zone load model, and the occupant's coordinates in the zone to approximate the thermal conditions near an occupant, which are then used for occupant comfort assessment. This model is integrated into the occupant comfort & behavior model as a MATLAB function prior to the estimation of occupant comfort.

## GEB Control Model

The purpose of the GEB Control Model is to generate supervisory control signals for the three testbeds in response to different GEB scenarios, i.e., energy efficiency, load shedding, load shifting, and load modulating. Using either rule-based control (RBC) or model predictive control (MPC), the GEB Control Model determines supervisory level setpoints, e.g., air handling unit (AHU) static pressure setpoint, to override system setpoints on the hardware side. More details about this model will be included in our future publications.

### Hardware Testbed

**IBAL HILFT.** This testbed utilizes the NIST IBAL facility (Pertzborn 2016; Pertzborn and Veronica 2018), which includes two chillers, one ice thermal storage tank, two AHUs, four variable air volume (VAV) units, one outdoor air emulator, and four zone emulators. The air-side system is a single-duct VAV with terminal reheat, which is typical of commercial and institutional buildings. Each of the two AHUs serves two VAV units (four total VAVs and zones), and both AHUs are served by the same primary-secondary hydronic system. The zone emulator is built into the duct system, with electric heater, steam spray humidifier, cooling coil, and thermal mass coil to emulate zone load.

**ASHP HILFT.** This testbed utilizes the NIST ASHP testing facility, which is equipped with two environmental chambers that emulate indoor and outdoor air conditions. Water-cooled, electrically heated AHUs are included in both chambers to create the outdoor weather conditions and the zone load.

**WSHP HILFT.** This testbed utilizes the TAMU WSHP facility (Calfa et al. 2022). This system is equipped with one zone emulator that is built into the duct system, and a hydronic system with a chiller and a water heater that provide constant flows of chilled or hot water as the heat sink or source to the WSHP and the air-side emulator load coil.

## INTEGRATION OF THE HARDWARE-IN-THE-LOOP TESTBED

### Integration Architecture

The HILFT integration is achieved through reliable co-simulation among various software environments including LabVIEW (Kalkman 1995), MATLAB (including Simulink), and EnergyPlus (via FMU). For all three testbeds, LabVIEW is the environment that controls the hardware and collects sensor measurements, and MATLAB is the environment that integrates all simulation activities. MongoDB (MongoDB 2018) is used for data storage, and data exchange among different simulation models. Figure 2 depicts the architecture of the HILFT integration

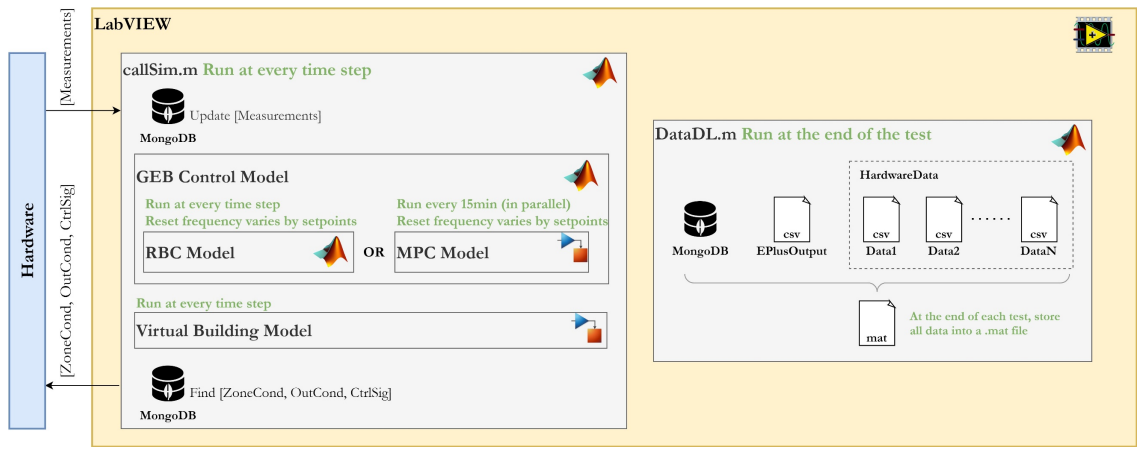

**Figure 2**  HILFT integration architecture

At each time step during a test, LabVIEW passes the collected sensor measurements that affect the Virtual Building Model and GEB Control Model to the software testbed (callSim.m) for real time communication. For example, the Virtual Building Model requires measurements of the discharge air temperature, relative humidity, and mass flow rate to the zone. The time step in this study is one minute. At every time step, the required measurements are stored (updated) in the MongoDB and are also used by the Virtual Building Model and the GEB Control Model to simulate the zone air conditions and control signals. To enable real-time simulation, i.e., force the model to execute in the same timeframe as the hardware, the Virtual Building Model (as a Simulink model) proceeds only one time step at each call, which can be achieved by using the "set_param" command in MATLAB to start/continue/pause the Simulink model.

After all variables have been simulated and stored within the current time step, callSim.m will retrieve (find) the simulated results from the database and pass them to LabVIEW. LabVIEW will then pass the control setpoints/schedules to the hardware equipment to operate the equipment accordingly.

At the end of each test, LabVIEW will call a MATLAB function (i.e., DataDL.m) to gather stored data from all places and store them into a .mat file. These data include the minute-by-minute data in the MongoDB, data in the EnergyPlus output file, and all data stored on the hardware side (including data that are collected by the hardware but not involved in the real time data exchange, such as fan speed, compressor speed, local control signals, etc.).

## Time Delay in the Integration

When integrating the HILFT hardware and software testbeds, time delays occur in various processes. Understanding the time delay is important when evaluating the quality of the integration. Overall, we have identified the following three types of delays: communication delay, control delay, and EnergyPlus inherited delay. Details of these delays are summarized below.

**Communication Delay (CommDelay).** Figure 3 illustrates how the hardware and software cooperates in real time. The communication delay is defined as the time between when the hardware passes measurement data to the simulation software and when the resulting updated simulation results are received by the hardware. This delay is contributed by the following three causes, the communication delay between LabVIEW and the hardware, the communication delay between LabVIEW and MATLAB, and the time consumed by the simulation within MATLAB. The delay of the first two is generally at the level of seconds or sub-seconds, while the time consumed by simulation varies with the size and complexity of the simulation problem. Excessive simulation time will prevent the software and hardware from communicating with each other properly. Our observation is that, for this study, the time needed for Virtual Building Model simulation is within a time step (one minute). When the GEB Control Model uses RBC strategies, its computation is within a time step too. Yet when the GEB Control Model uses MPC strategies, it might need more time than one time step. Therefore, in the test scenarios where MPC is used, the MPC model (as shown in Figure 2) is run in parallel to prevent long simulation time in real time communication.

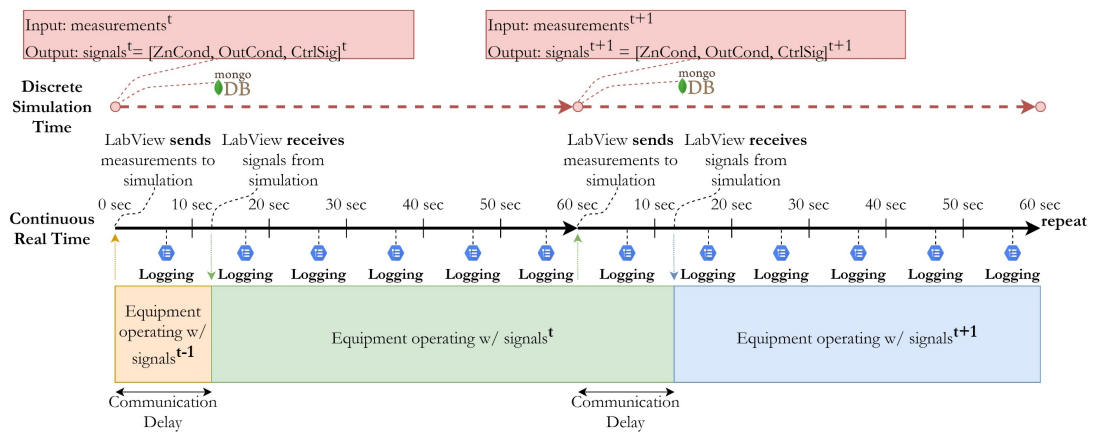

**Figure 3** Equipment real-time operation showing two time steps

**Control Delay.** To illustrate the control delay, Figure 4 depicts the data communication at a specific simulation time (e.g., 10:00 am). In this figure, first from left to right, the 10:00 discharge air condition is passed from the hardware side to the software side. Then, from right to left, the 10:00 zone air condition and outdoor air condition are passed back to the hardware side from the software side after the communication delay (CommDelay) as mentioned above. Due to the control delay, at this given moment, the 10:00 emulated results are the response to the 9:59 simulated results rather than the 10:00 simulated results. In other words, the simulated results at a given moment are always ahead of the emulated results. Therefore, when evaluating how well actual operating conditions of the hardware matches the corresponding values from the simulation, it is more reasonable to compare the emulated data of the current time step to the simulated data of the previous time step.

**EnergyPlus Inherited Delay**. During pre-testing we found that there is an inherited time delay on the software side due to the current solver structure of EnergyPlus 9.3. Even though the hardware data is successfully delivered to the software side at each simulation time step, EnergyPlus (as an FMU) does not use the current discharge air condition for the simulation, but rather the condition received at the previous time step due to its inherent setting, which results in a one simulation step delay on the simulation side. This delay is also shown in Figure 4, where the EnergyPlus discharge air condition at 10:00 is overridden by the 9:59 measurements instead of the 10:00 measurements. In other words, the control action of the HVAC system will not impact the simulated zone air until the next simulation time step. Considering that this project uses a 1-min time step, this delay is acceptable because the zone air dynamic is much slower in an actual building zone with typical thermal inertia. Further investigation is needed to eliminate this delay.

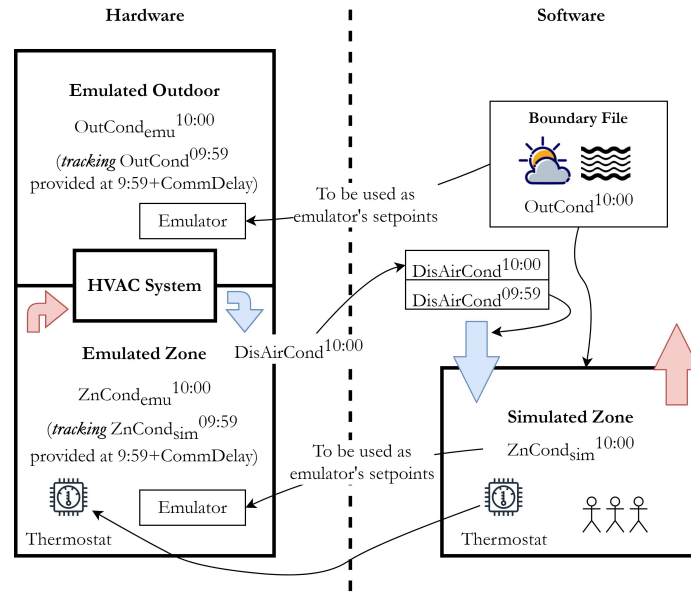

**Figure 4**   An example of data communication at simulation time 10:00

## Indoor/Outdoor Emulation

A successful integration relies on the ability of the indoor/outdoor emulators to accurately replicate the indoor/outdoor environment provided by the software side. For outdoor environment emulation, the control process is a straightforward feedback loop. In the IBAL, the outdoor air emulator uses the outdoor dry bulb temperature and outdoor relative humidity defined in the weather file as setpoints to generate the outdoor dry bulb temperature and outdoor relative humidity, respectively. ASHP's emulator does a similar job, except that it is the outdoor dew point temperature that is controlled instead of the outdoor relative humidity. For WSHP, the ground water temperature is the emulated variable.

The indoor environment emulation is more complex, especially for the IBAL and WSHP HILFTs, which have zone emulators built into the duct system. Figure 5 depicts the indoor air emulation. The zone emulator uses the simulated zone air condition as the setpoints to control the emulated zone air condition. These setpoints do not change during the one-minute communication interval. As shown in the figure, the HVAC controller at the hardware side can either use the emulated zone air condition, i.e., the real air condition of the emulated zone, or use the simulated zone air condition, i.e., simulated zone air temperature and humidity from the Virtual Building Model. For method 1, the process variable for HVAC system control is the emulated zone temperature. This control method is feasible in an ideal situation, where the emulated room temperature strictly follows the simulated room temperature, but in real situations there are control errors. If the internal mass of the emulated zone (i.e., any material in a zone that can capture and store heat to provide thermal inertia against temperature fluctuations) is insufficient, a strong coupling relationship will be formed between the emulated zone and the HVAC system. In this situation, control hunting will occur as the control error is amplified through the feedback loop. This control hunting issue was observed during the pre-testing of the IBAL and WSHP HILFTs. The main reason is that the zone air condition of the two testbeds is emulated using the heating/cooling coils and humidifier that are built into the duct system. Such emulated zones do not have sufficient internal mass. The ASHP testbed, on the other hand, has an indoor environmental chamber, which weakens the coupling effect between the emulated zone and the heat pump.

For an emulated zone with very light internal mass, such as those in the IBAL and the WSHP HILFTs, we recommend using method 2, where the process variable is the simulated zone temperature instead of the emulated zone temperature. After changing the process variable, the emulated zone and the HVAC system are decoupled, and a stable emulation was obtained in both testbeds.

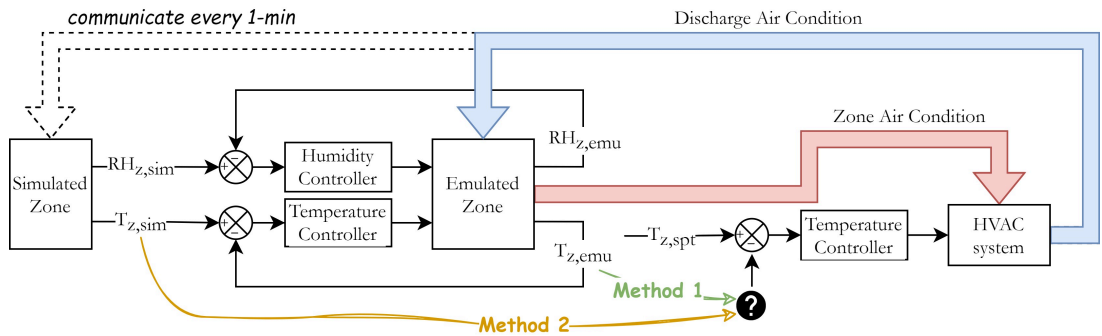

**Figure 5**  Indoor air emulation (method 2 is recommended for an emulated zone with light internal mass)

## Evaluating Integration Quality

To evaluate the integration quality, we have developed a set of key parameters which are presented below. Evaluating these parameters helped us identify integration issues during the pre-testing process. Given that these issues have not been fully addressed at the time of the writing of this article, we do not focus on the results in the following discussions, but rather on the purpose and definition of these parameters, and/or the methods used to measure them.

**Communication Delay.** The communication delay is defined as the time between when the hardware sending signals to the software side and when the resulting updated simulation results are received from the software side, as presented in Figure 3. This actual value of this delay cannot be determined accurately. However, an upper bound can be determined by comparing the data logged on the software side and the hardware side, because the measured and simulated data that are logged with the same time stamp on the software side do not appear at the same time on the hardware side. The upper bounds for communication time delays are 20 seconds, 25 seconds, and 5 seconds for the IBAL, ASHP, and WSHP HILFTs, respectively.

**HVAC System/Zone Emulator Capacity.** The equipment capacities were evaluated for the purpose of matching virtual zones with corresponding HVAC systems. The goal was to select simulated zones that are as diverse as possible, while ensuring that the peak load of the selected zones matches the rated capacities of the equipment.

**Outdoor Emulator Capacity.** For outdoor emulation, there is no special restriction for the WSHP and ASHP. WSHP's outdoor emulator can generate water temperature ranging from 10 °C (50 °F) to 55 °C (131 °F). ASHP's outdoor emulator can generate a variety of outdoor air conditions in the environmental chamber, with dry-bulb temperature ranging from -12 °C (10.4 °F) to 65 °C (149 °F) and relative humidity ranging from 10 % to 100 %. However, for the IBAL, considering that its outdoor air emulator is exposed to the local (NIST's location: Gaithersburg, MD) weather, there are limitations in generating certain weather conditions. For example, it is very difficult to emulate peak Atlanta, GA outdoor conditions during the real winter season at NIST. Therefore, a special test plan was developed for the IBAL based on the likelihood of generating certain weather conditions under the typical local weather at NIST.

**Outdoor/Indoor Emulation Response Time.** Using typical dynamic system response time definition (Skogestad 2008), the response times were measured by first bringing the system to a steady state (such as 70 °F and 50 % relative humidity) and then introducing a step change (such as changing the setpoint to be 72 °F). The response time (i.e., time constant) is then measured as the time from when the step change occurs until the system variable (i.e., air/water dry-bulb temperature, relative humidity/dewpoint temperature) reaches 63.2 % of the final change.

**Indoor/Outdoor Emulation Accuracy.** The accuracy of the emulation is evaluated using the root-mean-square-error (RMSE). Figure 6 (a) shows the comparison between simulated and emulated zone temperature in the ASHP HILFT. Considering the control delay (as shown in Figure 4), it is reasonable to compare simulated temperature to emulated temperature with a 1-min shift backward. The RMSEs for temperature comparison with and without the time shift are 0.06 °C (0.11 °F) and 0.07 °C (0.13 °F), respectively. Figure 6 (b) shows the comparison between simulated and emulated zone sensible cooling loads in the ASHP HILFT. The zone load is largely impacted by the discharge air condition. Considering the EnergyPlus inherent delay (as shown in Figure 4) in updating the discharge air condition, it is reasonable to compare simulated load to emulated loads with a 1-min shift forward. The RMSEs for load comparison with and without the time shift are 42.6 W and 560.5 W, respectively. The results show that, putting aside the time delay, the emulated and simulated data are essentially identical in shape.

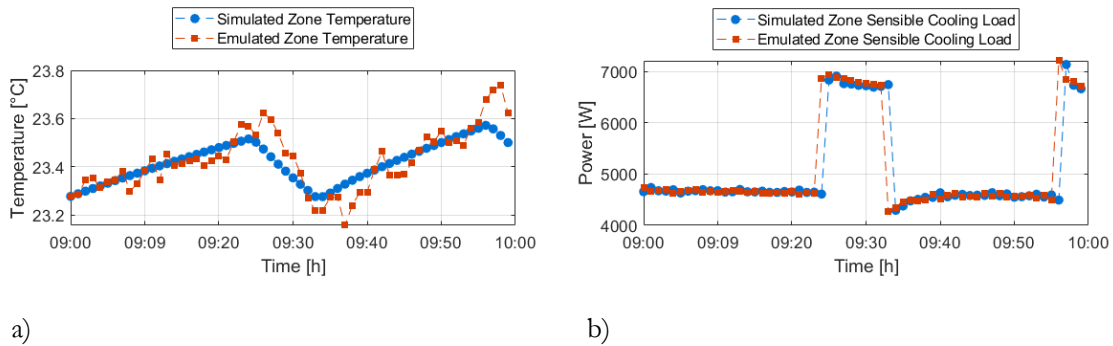

**Figure 6** (a) ASHP HILFT zone temperature and (b) ASHP HILFT zone sensible cooling load

## CONCLUSION

In this paper, three HILFTs that are capable of generating flexible load data for commercial buildings on a large scale are introduced. Important lessons learned during the integration of the software and hardware testbeds are discussed. Key integration parameters that are used to evaluate the effectiveness, robustness, and operational efficiency of the integration involving multiple devices and simulation programs are developed. Integration quality is greatly improved through the flexible use of databases, the careful design of the data structures, and the decoupling of the emulation control loop and the HVAC system control loop. Future studies will focus on analyzing the uncertainties of the testbeds and developing the confidence level of the load flexibility datasets.

## ACKNOWLEDGMENTS AND DISCLAIMER

This study is partially funded by the United States Department of Energy via grant EE-0009153. The views and opinions of authors expressed herein do not necessarily state or reflect those of the United States Government or any agency thereof.

## NOMENCLATURE

| | | |
|---|---|---|
| CtrlSig | = | control signals |
| DisAirCond | = | discharge air condition |
| OutCond | = | outdoor air condition |
| RHz | = | zone air relative humidity |
| Tz | = | zone air temperature |
| ZnCond | = | zone air condition |

**Subscripts**

| | | |
|---|---|---|
| *emu* | = | emulated |
| *sim* | = | simulated |
| *spt* | = | setpoint |